%% file: ga62_plb.tex
\documentclass{elsart}

\usepackage[isolatin]{inputenc}

\usepackage{graphicx}

\usepackage{amssymb}

\newcommand{\gall}{\nuc{62}{Ga}}
\newcommand{\zinc}{\nuc{62}{Zn}}
\newcommand{\nick}{\nuc{62}{Ni}}
\newcommand{\copp}{\nuc{62}{Cu}}
\newcommand{\result}[3]{ ($#1 \pm #2$)~#3}
\newcommand{\Ft}{$\mathcal{F}t$}
\newcommand{\zerotozero}{$0^+ \to 0^+$ }
\newcommand{\qec}{$Q_\mathrm{EC}$}
\begin{document}

\begin{frontmatter}


\title{Q-value of the superallowed $\beta$ decay of \gall}
\author[jyfl]{T. Eronen\corauthref{cor1}}
\ead{tommi.eronen@phys.jyu.fi},
\author[jyfl]{V. Elomaa},
\author[jyfl]{U. Hager},
\author[jyfl]{J. Hakala},
\author[jyfl]{A. Jokinen},
\author[jyfl]{I. Moore},
\author[jyfl]{H. Penttilä},
\author[jyfl]{S. Rahaman},
\author[jyfl]{S. Rinta-Antila},
\author[jyfl]{A. Saastamoinen},
\author[jyfl]{T. Sonoda} and
\author[jyfl]{J. Äystö}
\address[jyfl]{Department of Physics, P.O. Box 35 (YFL), FIN-40014 University of Jyväskylä, Finland}
\author[ganil]{A. Bey},
\author[ganil]{B. Blank},
\author[ganil]{G. Canchel},
\author[ganil]{C. Dossat},
\author[ganil]{J. Giovinazzo} and
\author[ganil]{I. Matea}


\corauth[cor1]{Corresponding author}

\address[ganil]{Centre d'études nucléaires de Bordeaux-Gradignan, Le Haut-Vigneau, F-33175 Gradignan Cedex, France}
\author[alger]{N. Adimi}
\address[alger]{Faculté de physique, USTHB, BP32, El Alia, 16111 Bab Ezzouar, Alger, Algeria}






\begin{abstract}
Masses of the radioactive isotopes \gall, \zinc{} and \copp{} have been measured at the JYFLTRAP facility 
with a relative precision of better than $1.8 \times 10^{-8}$. A $Q_\mathrm{EC}$ value of \result{9181.07}{0.54}{keV} for the superallowed decay of \gall{} is obtained from the measured cyclotron frequency ratios of
\gall---\zinc, \gall---\nick{} and \zinc---\nick{} ions. The resulting \Ft{}-value supports the validity of the conserved vector current hypothesis (CVC). The mass excess values measured were 
\result{-51986.5}{1.0}{keV} for \gall, 
\result{-61167.9}{0.9}{keV} for \zinc{}  and 
\result{-62787.2}{0.9}{keV} for \copp.

\end{abstract}

\end{frontmatter}

\section{Introduction}
\label{sec:intro}
According to the conserved vector current hypothesis (CVC), the matrix elements of the superallowed Fermi transitions between the $0^+$ isobaric analog states (IAS) should all be equal, independent of nuclear structure apart from small terms for radiative and isospin-symmetry breaking corrections. Provided this is true, the experimental values of the comparative half-life ($ft$) corrected for isospin mixing and radiative corrections \Ft, allow for an accurate determination of the weak vector coupling constant $G_\mathrm{V}$. This value combined 
with the muon decay constant $G_\mu$ allows the extraction of $V_\mathrm{ud}$, the weak 
coupling matrix element between up and down quarks in the  Cabibbo-Kobayashi-Maskawa (CKM) matrix. Combining this value with other data on the matrix elements of the first row of the CKM matrix, $V_\mathrm{us}$ and $V_\mathrm{ub}$, it becomes possible to 
test the unitarity of the CKM matrix and the validity of the electroweak 
standard model. Depending on which value for $V_\mathrm{us}$ is taken from the recent literature, the current world data as reviewed by Hardy and Towner implies a slight failure for the unitarity test by a maximum of 2.4 standard deviations \cite{har05}. More recently, a new measurement of the \qec-value of the superallowed $\beta$ decay of \nuc{46}{V} is indicating a need for the re-evaluation of the \qec-values of all the nine best-known cases employed in the unitarity analysis \cite{sav05}. Therefore, it is now becoming of utmost importance to improve the \qec-value data for these cases as well as to provide accurate data on new cases, as represented by \gall{} studied in this paper.

Following the notations of Hardy and Towner \cite{har05}, the true \Ft{}-value for the \zerotozero{} ($T=1$) decays is obtained using the equation
 
\begin{equation} \label{eq:ft}
\mathcal{F}t \equiv ft(1+\delta_\mathrm{R'})(1+\delta_\mathrm{NS}-\delta_\mathrm{C}) = \frac{K}{2G_\mathrm{V}^2\left (1+\Delta_\mathrm{R}^\mathrm{V} \right)} = \mathrm{constant},
\end{equation}

where $f$ is the $\beta$ decay phase-space factor, $t$ is the partial half-life, $K$ is a constant, $\delta_\mathrm{R'}$, $\delta_\mathrm{NS}$ and $\Delta_\mathrm{R} ^\mathrm{V}$ are radiative corrections and $\delta_\mathrm{C}$ accounts for the isospin-symmetry breaking correction, see ref. \cite{har05,tow02} for a detailed
discussion on these correction terms. A critical survey of 20 superallowed \zerotozero{} nuclear $\beta$ decays \cite{har05} shows that the corrected \Ft{}-values are constant to three parts in $10^4$. In the case of the subgroup of the nine best known \zerotozero{} transitions between $^{10}$C and $^{54}$Co the experimental $ft$-values are known to better than $0.15$\%.
Therefore, the correction terms ($\delta_\mathrm{NS}-\delta_\mathrm{C}$) and $\Delta_\mathrm{R}^\mathrm{V}$, obtained from theory, are becoming the limiting factor for improving the accuracy of the determination of the value of the vector coupling constant $G_\mathrm{V}$. Using equation (\ref{eq:ft}) together with the radiative corrections as determined by the theory and assuming the validity of the conserved vector current, the isospin correction $\delta_\mathrm{C}$ can be  experimentally determined providing thus important means to improve the theoretical model calculations. Currently, the dominant source of uncertainty in $\delta_\mathrm{C}$ arises from the disagreement between the results from different theoretical calculations.  

To complement the well-known cases, it is important to measure new cases where nuclear structure dependent corrections are large or particularly challenging for model calculations. Typically, these are nuclei with $T_z=-1$ as well as heavy $T_z=0$ nuclei with $Z>30$. Recently, three such cases --- $^{22}$Mg \cite{har03,muk04,sav04}, $^{34}$Ar \cite{her02} and $^{74}$Rb \cite{kel04,oin01,ba01} --- have been studied with high precision for their $Q_\mathrm{EC}$-value but with only a modest accuracy for the half-life or the superallowed decay branch resulting in more moderate accuracies for their \Ft-values ranging between 0.24 and 0.40 \% \cite{har05}.

The next heaviest nucleus in the series of $T_z=0$, $T=1$,  $I^\pi=0^+$ superallowed $\beta$ emitters beyond $^{54}$Co is \gall, which is a particularly interesting $T_z=0$ nucleus because its $\beta$ decay half-life has already been determined with high precision in several experiments yielding an average value of \result{116.175}{0.038}{ms} \cite{hym03,bla04,can05}. The current value and precision of the experimental branching ratio is ($99.85^{+0.05}_{-0.15}$)~\%.
However, possible additional unobserved Gamow-Teller transitions could increase the branching ratio to the excited $1^+$ states in \zinc, calling for more precise measurements of these transitions \cite{har02}. Prior to the present work, the $Q_\mathrm{EC}$-value of \gall{} has been determined in only one experiment where a $\beta$ end-point measurement resulted in a value of \result{9171}{26}{keV}. \cite{dav79}.

In this paper we present the first precise measurement of the $Q_\mathrm{EC}$-value and the mass of \gall{} employing a double Penning trap setup connected to the on-line isotope separator IGISOL at the University of Jyväskylä \cite{ays01}. In the same experiment, the masses of \zinc{} and \copp{} were determined using \nick{} as the reference ion whose mass excess is given with 0.6~keV uncertainty in the most recent Atomic Mass Evaluation tables \cite{aud03}.

\section{Experimental method}
\label{sec:method}
All nuclei of interest were produced in reactions induced by 48~MeV protons impinging on a 3 mg/cm$^2$ thick enriched $^{64}$Zn target. The recoil ions were slowed down and thermalized in the gas cell of an ion guide using 150~mbar helium pressure~\cite{hui04}. The ions were then transported by gas flow and electric fields through a differentially pumped electrode system into high vacuum and accelerated to 30~keV. After mass separation in a 55$^\circ$ dipole magnet the ions were injected into a buffer-gas filled RF-quadrupole for cooling and bunching before injection into a double Penning trap system for isobaric purification and mass measurements. The production rates with a 25~$\mu$A proton beam at the focal plane of IGISOL were measured at maximum to be 600 ions/s for \gall, $7\times 10^5$/s for \zinc, $10^6$/s for \copp{} and $10^5$/s for \nick. 

In a Penning trap ions are confined by a strong homogenous magnetic field and by an electric quadrupole field. The ions in the trap have three eigenmotions: one axial motion with a frequency of $\nu_z$ and two radial motions consisting of magnetron and reduced cyclotron motions  with
frequencies $\nu_-$ and $\nu_+$, respectively. The sum of these two frequencies equals to the true cyclotron frequency
\begin{equation} \label{eq:qbm}
\nu_- + \nu_+ = \nu_\mathrm{c} = \frac{1}{2\pi} \frac{qB}{m},
\end{equation}
where $q$ and $m$ are the charge and the mass of the ion and $B$ is the magnetic flux density \cite{bro86}.

The trap setup at JYFL, termed JYFLTRAP, consists of a gas-filled Radio Frequency Quadrupole Cooler (RFQ) \cite{nie02} and a double cylindrical Penning trap system which contains two traps: one for isobaric purification and the other for high precision mass measurements \cite{kol04}. The whole setup is placed on a high-voltage platform about 100~V below the level of the acceleration voltage of IGISOL. The two Penning traps are placed in the warm bore of a 7~Tesla superconducting magnet that has two homogeneous regions of 1~cm$^3$ in volume separated by 20 cm at the centre of the magnet. The purification trap has a field homogeneity $\frac{\Delta B}{B}$ of better than 10$^{-6}$ and the precision trap better than 10$^{-7}$. A mass resolving power $\frac{M}{\Delta M}$ of about 10$^5$ is typically obtained for the purification trap allowing for a clean separation of the studied $A=62$ isobars \cite{kol04,rin04} before injection to the precision trap. A mass measurement is performed by determining the true cyclotron frequency of the ion in the precision trap. The mass of the ion of interest is obtained by comparing its cyclotron frequency with a well known reference mass.

In a measurement cycle the ions are first captured and cooled in the buffer-gas filled RFQ which at a set time injects a low-energy ion bunch into the purification trap. In this trap the ions are captured by a time-varying electric potential, thermalized and cooled axially to the center of the potential  by collisions with helium buffer gas atoms. Then, the ions are excited by successive dipole and mass-selective quadrupole RF fields, centering only the ions of interest corresponding to their true cyclotron frequency given by the equation (\ref{eq:qbm}).

A typical cycle in the purification trap consisted of 25~ms axial cooling, 5~ms dipole (magnetron) and 40 ms quadrupole (cyclotron) excitation and finally 10~ms radial cooling before ejection into the precision trap which operates in vacuum. In the precision trap, the  magnetron radius of the ions is first increased with a dipole electric field for 5.5 ms using a phase-locking technique \cite{bla03} leading to a magnetron radius of about 0.6~mm. Then the quadrupole cyclotron excitation is switched on to mass-selectively convert the magnetron motion to reduced cyclotron motion. The duration of the quadrupole excitation was chosen to be between 100 and 400 ms and a corresponding amplitude to give rise to one full conversion in the resonance frequency of the equation (\ref{eq:qbm}). The detection of the resonance frequency is based on the time-of-flight technique \cite{kon95}. Due to large energy ratio (about 10$^6$) between the two radial motions, a significant decrease in the time-of-flight (TOF) from the trap to the ion detector is observed when the ions have even a modest component of reduced cyclotron motion. This is due to conversion of the radial energy into axial kinetic energy in the fringe field of the magnet. The absolute minimum of the time-of-flight occurs when the ions have been excited at their true cyclotron frequency.

\section{Results}
\label{sec:results}
The experimental setup was first tuned by using $^{40}$Ar$^+$ ions produced in collisions between the primary proton beam and argon atoms mixed in helium. Next, the $Q_\mathrm{EC}$-value of \gall{} was determined by measuring the cyclotron frequencies of \gall{} and \zinc{} in repeated successive measurements. The absolute mass values of \gall, \zinc{} and \copp{} were determined using \nick{} as the reference ion.
In the relative \qec{}-value measurements most of the systematic effects cancel out. The only contribution to the uncertainty arises from the measured frequency uncertainties, even without a contribution from the reference mass uncertainty.  The total error consists of statistical fitting uncertainty of a resonance curve as well as of systematic errors due to count rate variations and magnetic field fluctuations. In the absolute mass determination the uncertainty of the mass of the reference ion and possible mass-dependent errors have to be also included in the total error.

\begin{figure}[ht] \begin{center} 
\includegraphics{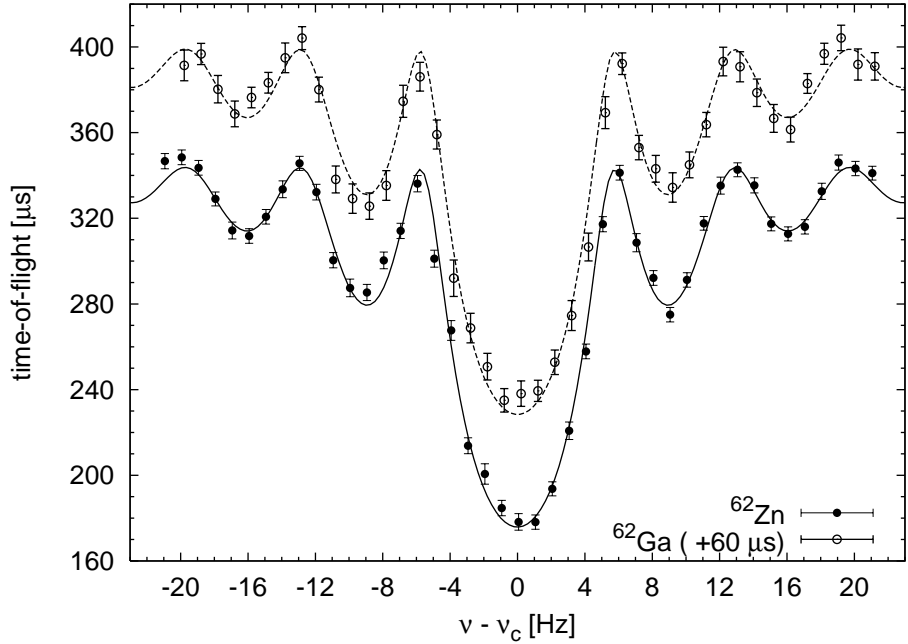}
\caption{Time-of-flight resonance curves for \gall{} and \zinc. For each frequency, 120 bunches were recorded 
to get the time-of-flight. An excitation time of 150~ms was used. The scanning time for one spectrum was typically 20~minutes and the
total number of ions detected about 5000. The time-of-flight resonance curve for \gall{} is offset by 60~$\mu$s for clarity.}
\label{fig:ga}
\end{center}\end{figure}

\begin{figure}[ht] \begin{center} 
\includegraphics{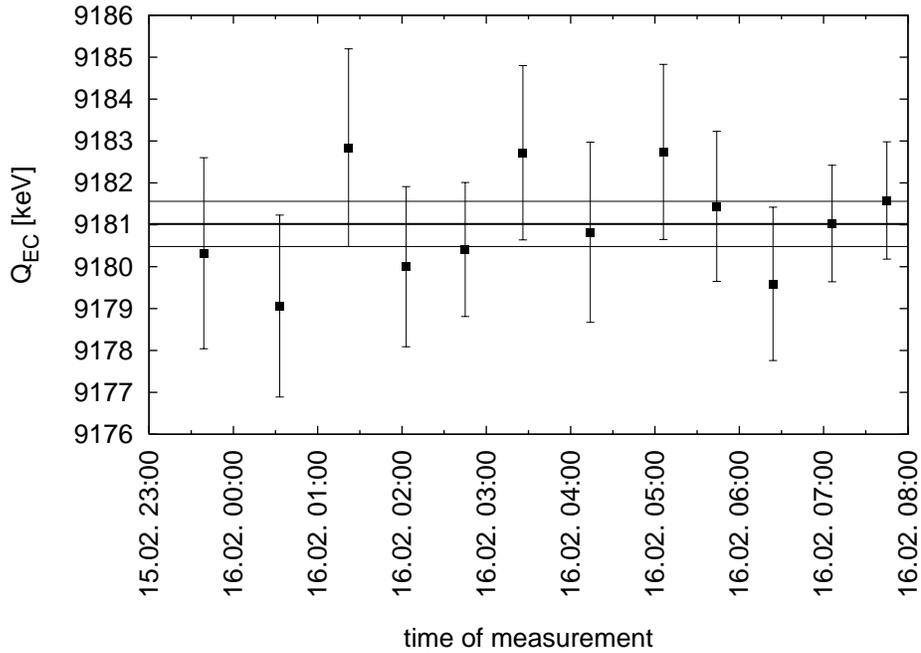}
\caption{A series of individual measurements of the \qec --value of \gall{} measured in february.
} \label{fig:collection}
\end{center}\end{figure}

Over the entire measurement period a total of nearly 50 resonances were recorded for both \gall{} and \zinc. In addition, the \qec-value of  \gall{}
was determined from several measurements of the \gall---\zinc{} and \zinc---\nick{} ion pairs.
Figure \ref{fig:ga} shows two samples of time-of-flight resonance curves measured for \gall{} and \zinc.
Figure \ref{fig:collection} shows the $Q_\mathrm{EC}$-values determined from 12 individual measurements with a 150~ms excitation time for both \gall{} and \zinc. Tables \ref{table:results1} and \ref{table:results2} summarize all the data obtained in all different experiments in February, March and September 2005.

The number of detected ions in each measurement cycle was kept below 2~ions/bunch in average. This was done to minimize the systematic uncertainty caused by high number of ions in the trap. At the time when the measurements were done, no individual recording of the ion bunches was possible. Later on, individual bunch recording was made possible and a count rate class \cite{kel03} analysis could be performed to characterise possible shifts in frequency due to a varying number of ions. Measurements of \copp{}  were done with 400~ms and 200~ms excitation times and varying the number of incoming ions. 
It was found that the true cyclotron frequency of \copp{} shifted to higher frequencies when the number of ions increased. The behaviour was found to be linear and the observed shift was determined to be 8 mHz /(1 ion/bunch).
A total shift of $\sqrt{2}\times 0.008$~Hz (corresponding to 0.38~keV) was taken as a systematic uncertainty to account for the shifts in both the reference ion and the ion of interest.
This compensates effects due to possible impurity ions in the trap originating, for instance, from radioactive decay of ions in the trap. 
The production rates of the measured ions slowly decreased while running the experiment. Even though the detection rate slowly decayed, no noticeable change in the obtained frequency ratios was observed. A small, slow fluctuation in the magnetic field strength was observed during the experiment. This, however, had a negligible effect on the final uncertainty  due to a large number of reference measurements made.

The final \qec{} value for the \gall{} decay obtained as a weighted mean of the relative measurements given in table \ref{table:results1} is \result{9181.07}{0.38}{keV}. Here the uncertainty is derived from the statistical uncertainties of the resonance frequencies measured for these two ions. 
A total error of 0.54~keV is obtained by adding the statistical (0.38~keV) and systematic (0.38~keV) errors quadratically.

The absolute masses of \gall{} and \zinc{} were determined using \nick{} as a reference ion.
These measurements resulted in the mass excess values of \result{-51986.5}{1.0}{keV} and 
\result{-61167.9}{0.9}{keV} for \gall{} and \zinc, respectively, when using the mass excess value of
\result{-66746.1}{0.6}{keV} for the reference ion \nick{} \cite{aud03}.
Note that in addition to the statistical and systematic errors, the mass excess values have an error contribution also from 
the reference mass uncertainty. 
The new mass excess value of \zinc{} is in agreement with but
more precise than the previously adopted value of \result{-61171}{10}{keV}. The
mass excess of \copp{} was also measured. The obtained value of \result{-62787.2}{0.9}{keV} deviates
from the previously adopted value of \result{-62798}{4}{keV} given in \cite{aud03}. However, the previous
experimental values obtained via $\beta$ end-point and the ($p$,$n$) reaction threshold
measurements scatter by as much as 24~keV with individual uncertainties of 10~keV. 

\section{Discussion}
\label{sec:discussion}
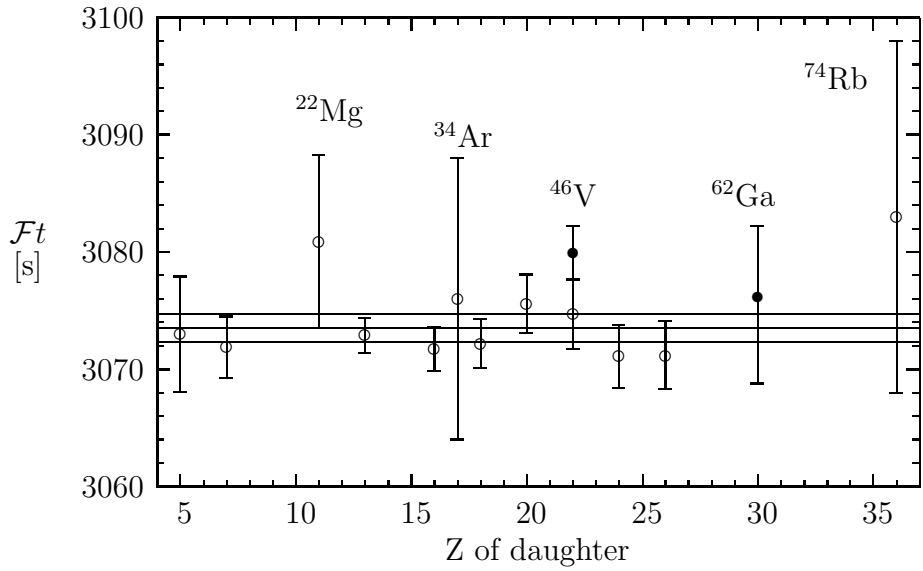
\begin{figure}[ht] \begin{center} 
\input{ft_values.tex}
\caption{ \Ft-values for the most precisely known superallowed $\beta$ emitters. The five most recently measured cases are labelled by the parent nucleus. The horizontal lines denote one standard deviation around the world average \Ft-value of \result{3073.5}{1.2} given in ref. \cite{har05}. The newest values marked by filled circles for \nuc{46}{V} and \gall{} are not included in this average.}
\label{fig:ft_values}
\end{center}\end{figure}

The new experimental value of \result{9181.07}{0.54}{keV}  for the decay energy of \gall{} yields a value of \result{26401.6}{8.3} for the statistical rate function $f$ \cite{har05,har05_private}. The half-life of \result{116.175}{0.038}{ms} and the branching ratio of (99.85$^{+0.05}_{-0.15}$)\%
yield the final  $ft$-value of (3076.0$^{+2.1}_{-4.8}$)~s for \gall. This value, when corrected for radiative and isospin mixing effects according to Table IX of ref. \cite{har05}, results in an \Ft-value of 
($3076.2^{+6.0}_{-7.4}$) which is in good agreement with the current world average of \result{3073.5}{1.2}  given in the review by Hardy and Towner \cite{har05}, see also figure \ref{fig:ft_values}. The experimental contribution to the uncertainty is now dominated by the branching ratio measurement. The overall uncertainty, however, is strongly influenced by the uncertainties of the theoretical corrections, especially of ($1-\delta_\mathrm{C}$) whose relative uncertainty is as large as $1.6\times 10^{-3}$. 

In order to gain a deeper understanding of the calculation of $\delta_\mathrm{C}$ it is of interest to solve the experimental value of $\delta_\mathrm{C}$ employing the CVC in the framework of equation (\ref{eq:ft}). Using a value of $\delta_\mathrm{NS}$ from \cite{tow02},  $\delta_\mathrm{C}$~=~($1.47^{+0.12}_{-0.18}$) is obtained which is somewhat higher but still in agreement with the adopted theoretical value of \result{1.38}{0.16} given in \cite{tow02} as well as with the value of 1.42 obtained by Sagawa et al. \cite{sag96}. In the paper of Ormand and Brown the value of $\delta_\mathrm{C}$ varies between 1.30 and 1.69,  respectively, when Hartree-Fock or Woods-Saxon (WS) wave functions are used \cite{orm95}. Therefore, our result does not favor either of these two calculations.

In conclusion, the \qec-value for the superallowed decay of \gall{} has been measured with a precision comparable to the nine best known cases. The corresponding $ft$-value is more precise than the other recently measured ``new'' cases, \nuc{22}{Mg}, \nuc{34}{Ar} and \nuc{74}{Rb}. Therefore, it has become of great importance to measure the branching ratio more precisely in order that the decay of \gall{} becomes relevant as a contribution to the unitarity test of the CKM matrix.
As pointed out earlier, a new measurement of the \qec-value of the superallowed decay of \nuc{46}{V} is indicating a need for the re-evaluation of the \qec-values of all of the nine best-known cases \cite{sav05}. Currently, the \Ft-value for \nuc{46}{V} is the only one outside of one standard deviation of the current world average value. 
This necessitates new measurements for the \qec-values of the well-known cases such as \nuc{26}{Al}, \nuc{42}{Sc}, \nuc{46}{V}, \nuc{50}{Mn} and \nuc{54}{Co} which all are available as ion beams at the JYFLTRAP facility.

\input{acknowledgements.tex}



\bibliographystyle{elsart-num}
\bibliography{PRLrefs_long}
\newpage
\begin{table}[ht]
\caption{Summary of the data from the \qec-value measurements. The uncertainties given are statistical and due to the fitting procedure of the TOF-frequency curves only. The symbol \# denotes the number of measurements done. ($^\dagger$) \qec-value determined by using the frequency ratios of \gall--\nick{} and \zinc--\nick{} from table \ref{table:results2}.  }
\begin{tabular}[htp]{|c|c|c|c|c|c|l|l|c|}
\hline
&measurement/			&	$T_\mathrm{RF}$	&	\#	&	frequency	&	$Q_\mathrm{EC}$	\\
&reference			&		[ms]		&		&	ratio		&	[keV]			\\ \hline
feb	&	\gall	/	\zinc	&	150	&	12	&	1.0001591414(92)$\phantom{0}$ &	9181.02(53) \\
mar	&	\gall	/	\zinc	&	150	&	27	&	1.0001591337(135) &	9180.57(78) \\
mar	&	\gall	/	\zinc	&	100	&	7	&	1.0001591692(274) &	9182.62(158) \\
Ga-Zn/Ni $^\dagger$&	\nick{} as ref		&	150/400	&		&		&	9181.34(86) \\ \hline
average &				&		&		&			&	9181.07(38) \\ \hline 
\end{tabular}

\label{table:results1}
\end{table}


\begin{table}[ht]
\caption{Summary of the mass excess values obtained by using \nick{} as the reference ion. The uncertainties given are statistical and due to the fitting procedure of the TOF-frequency curves only. The symbol \# denotes the number of measurements done.}
\begin{tabular}[htp]{|c|c|c|c|c|c|c|l|c|}
\hline
&measurement/			&	$T_\mathrm{RF}$	&	\#	&	frequency	&	mass excess 	\\
&reference			&		[ms]		&		&	ratio		&	[keV]			\\ \hline
feb	&	\gall	/	\nick	&	150	&	5	&	1.0002558705(157) &				\\
mar	&				&	100	&	10	&	1.0002558552(217) &	 \\
mar	&				&	150	&	7	&	1.0002558554(271) &				\\ \hline
average &				&		&		&	1.0002558635(115) &	-51986.52(66) \\ \hline \hline
feb	&	\zinc	/	\nick	&	400	&	4	&	1.0000966941(123) &				\\
sep	&				&	400	&	3	&	1.0000967109(145) &	 \\ \hline
average &				&		&		&	1.0000967012(94)$\phantom{0}$ &	-61167.86(54) \\ \hline \hline
feb	&	\copp	/	\nick	&	400	&	1	&	1.0000686297(235) &				\\
sep	&				&	400	&	7	&	1.0000686290(87)$\phantom{0}$ &	 \\ \hline
average &				&		&		&	1.0000686291(82)$\phantom{0}$ &	 -62787.20(47)\\ \hline

\end{tabular}
\label{table:results2}
\end{table}

\end{document}

%% file: acknowledgements.tex
\section{Acknowledgements}
\label{sec:acknowledgements}
This work has been supported by the EU within the 6th framework programme Integrating
Infrastructure Initiative - Transnational Access, Contract Number:506065
(EURONS) and within the NIPNET RTD project under Contract No. HPRI-CT-2001-50034.
We also acknowledge support from 
the Academy of Finland under the Finnish Centre of Excellence
Programme 2000-2005 (Project No. 44875, Nuclear and Condensed Matter Physics
Programme at JYFL) and the Conseil Régional d'Aquitaine. 
A.J. and H.P. are indebted to financial support from the Academy of Finland 
(Project numbers 46351 and 202256).

%% file: ft_values.tex
\setlength{\unitlength}{0.240900pt}
\ifx\plotpoint\undefined\newsavebox{\plotpoint}\fi
\sbox{\plotpoint}{\rule[-0.200pt]{0.400pt}{0.400pt}}%
\begin{picture}(1500,900)(0,0)
\sbox{\plotpoint}{\rule[-0.200pt]{0.400pt}{0.400pt}}%
\put(242.0,123.0){\rule[-0.200pt]{4.818pt}{0.400pt}}
\put(222,123){\makebox(0,0)[r]{ 3060}}
\put(1419.0,123.0){\rule[-0.200pt]{4.818pt}{0.400pt}}
\put(242.0,160.0){\rule[-0.200pt]{2.409pt}{0.400pt}}
\put(1429.0,160.0){\rule[-0.200pt]{2.409pt}{0.400pt}}
\put(242.0,197.0){\rule[-0.200pt]{2.409pt}{0.400pt}}
\put(1429.0,197.0){\rule[-0.200pt]{2.409pt}{0.400pt}}
\put(242.0,234.0){\rule[-0.200pt]{2.409pt}{0.400pt}}
\put(1429.0,234.0){\rule[-0.200pt]{2.409pt}{0.400pt}}
\put(242.0,270.0){\rule[-0.200pt]{2.409pt}{0.400pt}}
\put(1429.0,270.0){\rule[-0.200pt]{2.409pt}{0.400pt}}
\put(242.0,307.0){\rule[-0.200pt]{4.818pt}{0.400pt}}
\put(222,307){\makebox(0,0)[r]{ 3070}}
\put(1419.0,307.0){\rule[-0.200pt]{4.818pt}{0.400pt}}
\put(242.0,344.0){\rule[-0.200pt]{2.409pt}{0.400pt}}
\put(1429.0,344.0){\rule[-0.200pt]{2.409pt}{0.400pt}}
\put(242.0,381.0){\rule[-0.200pt]{2.409pt}{0.400pt}}
\put(1429.0,381.0){\rule[-0.200pt]{2.409pt}{0.400pt}}
\put(242.0,418.0){\rule[-0.200pt]{2.409pt}{0.400pt}}
\put(1429.0,418.0){\rule[-0.200pt]{2.409pt}{0.400pt}}
\put(242.0,455.0){\rule[-0.200pt]{2.409pt}{0.400pt}}
\put(1429.0,455.0){\rule[-0.200pt]{2.409pt}{0.400pt}}
\put(242.0,492.0){\rule[-0.200pt]{4.818pt}{0.400pt}}
\put(222,492){\makebox(0,0)[r]{ 3080}}
\put(1419.0,492.0){\rule[-0.200pt]{4.818pt}{0.400pt}}
\put(242.0,528.0){\rule[-0.200pt]{2.409pt}{0.400pt}}
\put(1429.0,528.0){\rule[-0.200pt]{2.409pt}{0.400pt}}
\put(242.0,565.0){\rule[-0.200pt]{2.409pt}{0.400pt}}
\put(1429.0,565.0){\rule[-0.200pt]{2.409pt}{0.400pt}}
\put(242.0,602.0){\rule[-0.200pt]{2.409pt}{0.400pt}}
\put(1429.0,602.0){\rule[-0.200pt]{2.409pt}{0.400pt}}
\put(242.0,639.0){\rule[-0.200pt]{2.409pt}{0.400pt}}
\put(1429.0,639.0){\rule[-0.200pt]{2.409pt}{0.400pt}}
\put(242.0,676.0){\rule[-0.200pt]{4.818pt}{0.400pt}}
\put(222,676){\makebox(0,0)[r]{ 3090}}
\put(1419.0,676.0){\rule[-0.200pt]{4.818pt}{0.400pt}}
\put(242.0,713.0){\rule[-0.200pt]{2.409pt}{0.400pt}}
\put(1429.0,713.0){\rule[-0.200pt]{2.409pt}{0.400pt}}
\put(242.0,749.0){\rule[-0.200pt]{2.409pt}{0.400pt}}
\put(1429.0,749.0){\rule[-0.200pt]{2.409pt}{0.400pt}}
\put(242.0,786.0){\rule[-0.200pt]{2.409pt}{0.400pt}}
\put(1429.0,786.0){\rule[-0.200pt]{2.409pt}{0.400pt}}
\put(242.0,823.0){\rule[-0.200pt]{2.409pt}{0.400pt}}
\put(1429.0,823.0){\rule[-0.200pt]{2.409pt}{0.400pt}}
\put(242.0,860.0){\rule[-0.200pt]{4.818pt}{0.400pt}}
\put(222,860){\makebox(0,0)[r]{ 3100}}
\put(1419.0,860.0){\rule[-0.200pt]{4.818pt}{0.400pt}}
\put(242.0,123.0){\rule[-0.200pt]{0.400pt}{2.409pt}}
\put(242.0,850.0){\rule[-0.200pt]{0.400pt}{2.409pt}}
\put(278.0,123.0){\rule[-0.200pt]{0.400pt}{4.818pt}}
\put(278,82){\makebox(0,0){ 5}}
\put(278.0,840.0){\rule[-0.200pt]{0.400pt}{4.818pt}}
\put(315.0,123.0){\rule[-0.200pt]{0.400pt}{2.409pt}}
\put(315.0,850.0){\rule[-0.200pt]{0.400pt}{2.409pt}}
\put(351.0,123.0){\rule[-0.200pt]{0.400pt}{2.409pt}}
\put(351.0,850.0){\rule[-0.200pt]{0.400pt}{2.409pt}}
\put(387.0,123.0){\rule[-0.200pt]{0.400pt}{2.409pt}}
\put(387.0,850.0){\rule[-0.200pt]{0.400pt}{2.409pt}}
\put(423.0,123.0){\rule[-0.200pt]{0.400pt}{2.409pt}}
\put(423.0,850.0){\rule[-0.200pt]{0.400pt}{2.409pt}}
\put(460.0,123.0){\rule[-0.200pt]{0.400pt}{4.818pt}}
\put(460,82){\makebox(0,0){ 10}}
\put(460.0,840.0){\rule[-0.200pt]{0.400pt}{4.818pt}}
\put(496.0,123.0){\rule[-0.200pt]{0.400pt}{2.409pt}}
\put(496.0,850.0){\rule[-0.200pt]{0.400pt}{2.409pt}}
\put(532.0,123.0){\rule[-0.200pt]{0.400pt}{2.409pt}}
\put(532.0,850.0){\rule[-0.200pt]{0.400pt}{2.409pt}}
\put(568.0,123.0){\rule[-0.200pt]{0.400pt}{2.409pt}}
\put(568.0,850.0){\rule[-0.200pt]{0.400pt}{2.409pt}}
\put(605.0,123.0){\rule[-0.200pt]{0.400pt}{2.409pt}}
\put(605.0,850.0){\rule[-0.200pt]{0.400pt}{2.409pt}}
\put(641.0,123.0){\rule[-0.200pt]{0.400pt}{4.818pt}}
\put(641,82){\makebox(0,0){ 15}}
\put(641.0,840.0){\rule[-0.200pt]{0.400pt}{4.818pt}}
\put(677.0,123.0){\rule[-0.200pt]{0.400pt}{2.409pt}}
\put(677.0,850.0){\rule[-0.200pt]{0.400pt}{2.409pt}}
\put(714.0,123.0){\rule[-0.200pt]{0.400pt}{2.409pt}}
\put(714.0,850.0){\rule[-0.200pt]{0.400pt}{2.409pt}}
\put(750.0,123.0){\rule[-0.200pt]{0.400pt}{2.409pt}}
\put(750.0,850.0){\rule[-0.200pt]{0.400pt}{2.409pt}}
\put(786.0,123.0){\rule[-0.200pt]{0.400pt}{2.409pt}}
\put(786.0,850.0){\rule[-0.200pt]{0.400pt}{2.409pt}}
\put(822.0,123.0){\rule[-0.200pt]{0.400pt}{4.818pt}}
\put(822,82){\makebox(0,0){ 20}}
\put(822.0,840.0){\rule[-0.200pt]{0.400pt}{4.818pt}}
\put(859.0,123.0){\rule[-0.200pt]{0.400pt}{2.409pt}}
\put(859.0,850.0){\rule[-0.200pt]{0.400pt}{2.409pt}}
\put(895.0,123.0){\rule[-0.200pt]{0.400pt}{2.409pt}}
\put(895.0,850.0){\rule[-0.200pt]{0.400pt}{2.409pt}}
\put(931.0,123.0){\rule[-0.200pt]{0.400pt}{2.409pt}}
\put(931.0,850.0){\rule[-0.200pt]{0.400pt}{2.409pt}}
\put(967.0,123.0){\rule[-0.200pt]{0.400pt}{2.409pt}}
\put(967.0,850.0){\rule[-0.200pt]{0.400pt}{2.409pt}}
\put(1004.0,123.0){\rule[-0.200pt]{0.400pt}{4.818pt}}
\put(1004,82){\makebox(0,0){ 25}}
\put(1004.0,840.0){\rule[-0.200pt]{0.400pt}{4.818pt}}
\put(1040.0,123.0){\rule[-0.200pt]{0.400pt}{2.409pt}}
\put(1040.0,850.0){\rule[-0.200pt]{0.400pt}{2.409pt}}
\put(1076.0,123.0){\rule[-0.200pt]{0.400pt}{2.409pt}}
\put(1076.0,850.0){\rule[-0.200pt]{0.400pt}{2.409pt}}
\put(1113.0,123.0){\rule[-0.200pt]{0.400pt}{2.409pt}}
\put(1113.0,850.0){\rule[-0.200pt]{0.400pt}{2.409pt}}
\put(1149.0,123.0){\rule[-0.200pt]{0.400pt}{2.409pt}}
\put(1149.0,850.0){\rule[-0.200pt]{0.400pt}{2.409pt}}
\put(1185.0,123.0){\rule[-0.200pt]{0.400pt}{4.818pt}}
\put(1185,82){\makebox(0,0){ 30}}
\put(1185.0,840.0){\rule[-0.200pt]{0.400pt}{4.818pt}}
\put(1221.0,123.0){\rule[-0.200pt]{0.400pt}{2.409pt}}
\put(1221.0,850.0){\rule[-0.200pt]{0.400pt}{2.409pt}}
\put(1258.0,123.0){\rule[-0.200pt]{0.400pt}{2.409pt}}
\put(1258.0,850.0){\rule[-0.200pt]{0.400pt}{2.409pt}}
\put(1294.0,123.0){\rule[-0.200pt]{0.400pt}{2.409pt}}
\put(1294.0,850.0){\rule[-0.200pt]{0.400pt}{2.409pt}}
\put(1330.0,123.0){\rule[-0.200pt]{0.400pt}{2.409pt}}
\put(1330.0,850.0){\rule[-0.200pt]{0.400pt}{2.409pt}}
\put(1366.0,123.0){\rule[-0.200pt]{0.400pt}{4.818pt}}
\put(1366,82){\makebox(0,0){ 35}}
\put(1366.0,840.0){\rule[-0.200pt]{0.400pt}{4.818pt}}
\put(1403.0,123.0){\rule[-0.200pt]{0.400pt}{2.409pt}}
\put(1403.0,850.0){\rule[-0.200pt]{0.400pt}{2.409pt}}
\put(1439.0,123.0){\rule[-0.200pt]{0.400pt}{2.409pt}}
\put(1439.0,850.0){\rule[-0.200pt]{0.400pt}{2.409pt}}
\put(242.0,123.0){\rule[-0.200pt]{288.357pt}{0.400pt}}
\put(1439.0,123.0){\rule[-0.200pt]{0.400pt}{177.543pt}}
\put(242.0,860.0){\rule[-0.200pt]{288.357pt}{0.400pt}}
\put(242.0,123.0){\rule[-0.200pt]{0.400pt}{177.543pt}}
\put(40,491){\makebox(0,0){\shortstack{$\mathcal{F}t$\\ $[$s$]$}}}
\put(840,21){\makebox(0,0){Z of daughter}}
\put(460,713){\makebox(0,0)[l]{$^{22}$Mg}}
\put(677,676){\makebox(0,0)[l]{$^{34}$Ar}}
\put(859,584){\makebox(0,0)[l]{$^{46}$V}}
\put(1113,584){\makebox(0,0)[l]{$^{62}$Ga}}
\put(1258,768){\makebox(0,0)[l]{$^{74}$Rb}}
\put(242,372){\line(1,0){1197}}
\put(242,350){\line(1,0){1197}}
\put(242,394){\line(1,0){1197}}
\put(278.0,272.0){\rule[-0.200pt]{0.400pt}{43.603pt}}
\put(268.0,272.0){\rule[-0.200pt]{4.818pt}{0.400pt}}
\put(268.0,453.0){\rule[-0.200pt]{4.818pt}{0.400pt}}
\put(351.0,294.0){\rule[-0.200pt]{0.400pt}{23.126pt}}
\put(341.0,294.0){\rule[-0.200pt]{4.818pt}{0.400pt}}
\put(341.0,390.0){\rule[-0.200pt]{4.818pt}{0.400pt}}
\put(496.0,372.0){\rule[-0.200pt]{0.400pt}{65.525pt}}
\put(486.0,372.0){\rule[-0.200pt]{4.818pt}{0.400pt}}
\put(486.0,644.0){\rule[-0.200pt]{4.818pt}{0.400pt}}
\put(568.0,333.0){\rule[-0.200pt]{0.400pt}{13.249pt}}
\put(558.0,333.0){\rule[-0.200pt]{4.818pt}{0.400pt}}
\put(558.0,388.0){\rule[-0.200pt]{4.818pt}{0.400pt}}
\put(677.0,304.0){\rule[-0.200pt]{0.400pt}{16.863pt}}
\put(667.0,304.0){\rule[-0.200pt]{4.818pt}{0.400pt}}
\put(667.0,374.0){\rule[-0.200pt]{4.818pt}{0.400pt}}
\put(714.0,197.0){\rule[-0.200pt]{0.400pt}{106.478pt}}
\put(704.0,197.0){\rule[-0.200pt]{4.818pt}{0.400pt}}
\put(704.0,639.0){\rule[-0.200pt]{4.818pt}{0.400pt}}
\put(750.0,309.0){\rule[-0.200pt]{0.400pt}{18.549pt}}
\put(740.0,309.0){\rule[-0.200pt]{4.818pt}{0.400pt}}
\put(740.0,386.0){\rule[-0.200pt]{4.818pt}{0.400pt}}
\put(822.0,364.0){\rule[-0.200pt]{0.400pt}{22.163pt}}
\put(812.0,364.0){\rule[-0.200pt]{4.818pt}{0.400pt}}
\put(812.0,456.0){\rule[-0.200pt]{4.818pt}{0.400pt}}
\put(895.0,339.0){\rule[-0.200pt]{0.400pt}{26.499pt}}
\put(885.0,339.0){\rule[-0.200pt]{4.818pt}{0.400pt}}
\put(885.0,449.0){\rule[-0.200pt]{4.818pt}{0.400pt}}
\put(967.0,278.0){\rule[-0.200pt]{0.400pt}{23.849pt}}
\put(957.0,278.0){\rule[-0.200pt]{4.818pt}{0.400pt}}
\put(957.0,377.0){\rule[-0.200pt]{4.818pt}{0.400pt}}
\put(1040.0,276.0){\rule[-0.200pt]{0.400pt}{25.776pt}}
\put(1030.0,276.0){\rule[-0.200pt]{4.818pt}{0.400pt}}
\put(1030.0,383.0){\rule[-0.200pt]{4.818pt}{0.400pt}}
\put(1403.0,270.0){\rule[-0.200pt]{0.400pt}{133.218pt}}
\put(1393.0,270.0){\rule[-0.200pt]{4.818pt}{0.400pt}}
\put(278,363){\circle{18}}
\put(351,342){\circle{18}}
\put(496,508){\circle{18}}
\put(568,361){\circle{18}}
\put(677,339){\circle{18}}
\put(714,418){\circle{18}}
\put(750,348){\circle{18}}
\put(822,410){\circle{18}}
\put(895,394){\circle{18}}
\put(967,328){\circle{18}}
\put(1040,329){\circle{18}}
\put(1403,547){\circle{18}}
\put(1393.0,823.0){\rule[-0.200pt]{4.818pt}{0.400pt}}
\put(895.0,447.0){\rule[-0.200pt]{0.400pt}{20.476pt}}
\put(885.0,447.0){\rule[-0.200pt]{4.818pt}{0.400pt}}
\put(895,490){\circle*{18}}
\put(885.0,532.0){\rule[-0.200pt]{4.818pt}{0.400pt}}
\put(1185.0,285.0){\rule[-0.200pt]{0.400pt}{59.502pt}}
\put(1175.0,532.0){\rule[-0.200pt]{4.818pt}{0.400pt}}
\put(1185,421){\circle*{18}}
\put(1175.0,285.0){\rule[-0.200pt]{4.818pt}{0.400pt}}
\put(242.0,123.0){\rule[-0.200pt]{288.357pt}{0.400pt}}
\put(1439.0,123.0){\rule[-0.200pt]{0.400pt}{177.543pt}}
\put(242.0,860.0){\rule[-0.200pt]{288.357pt}{0.400pt}}
\put(242.0,123.0){\rule[-0.200pt]{0.400pt}{177.543pt}}
\end{picture}